# Luhmann Reconsidered:
# Steps Towards an Empirical Research Programme in the Sociology of Communication?

Loet Leydesdorff
forthcoming in: Colin Grant (ed.), *Beyond Universal Pragmatics: Essays in the Philosophy of Communication.* Oxford: Peter Lang.

It was not without irony that the philosopher Karl Popper (1963, at p. 35) told the following anecdote about a meeting with the psychoanalyst Alfred Adler:

> Once, in 1919, I reported to him a case which to me did not seem particularly Adlerian, but which he found no difficulty in analyzing in terms of his theory of inferiority feelings, although he had not even seen the child. Slightly shocked, I asked him how he could be so sure. 'Because of my thousandfold experience,' he replied; whereupon I could not help saying: 'And with this new case, I suppose, your experience has become thousand-and-one fold.'

Although Luhmann formulated with modesty and precaution, for example in *Die Wissenschaft der Gesellschaft* (1990a, at pp. 412f.), that his theory claims to be a universal one because it is self-referential, the "operational closure" that follows from this assumption easily generates a problem for empirical research. Can a theory which considers society—and science as one of its subsystems—operationally closed, nevertheless contribute to the project of Enlightenment which Popper (1945) so vigorously identified as the driver of an *open* society? How can a theory which proclaims itself to be circular and universal nevertheless claim to celebrate "the triumph of the Enlightenment" (Luhmann, 1990a, at p. 548)? Is the lack of an empirical program of research building on Luhmann's theory fortuitous or does it indicate that this theory should be considered as a philosophy rather than a *heuristic* for the explanation of operations in social systems?

In my opinion, Luhmann's sociological theory of communications contains important elements which have hitherto not sufficiently been appreciated in the empirical traditions of sociology and communication studies (Leydesdorff, 1996; Seidl & Becker, 2006; Grant, 2007). Anthony Giddens (1984, at p. xxxvii), for example, had no doubt that "these newer versions of Parsonianism, particularly Luhmann and Habermas, were to be repudiated despite the sophistication and importance of these authors." However, Giddens focused on explaining action; social structure was black-boxed in his "structuration theory" as a "duality" which precedes action as "rules and resources," and follows from the aggregation of human actions, for example, as institutions (Leydesdorff, 1993). According to Giddens (1984), social structures exist in social reality only by implication, i.e., in their "instantiation" in the knowledgeable activities of situated actors. This duality of social structure cannot be specified empirically without reference to actions and institutions because structure is considered "outside of time-space" (Giddens, 1981, at pp. 171f.) and as an "absent set of differences" (Giddens, 1979, at p. 64).

Giddens's "virtuality" of structure can also be considered as a dynamic extension of the sociological concept of latency (Lazersfeld & Henry, 1968): the structural dimensions of a social network system are not manifest to participating agents. The agents may be able to conjecture these dimensions reflexively, but predictably to a variable extent. However, Luhmann (1984) theorized about social systems of communication as structural, yet not directly observable dynamics;[1] human agents ("consciousness") were defined as the (structurally coupled and therefore necessary) environment of systems of social coordination (Luhmann, 1984, 1986a, 2002). Nevertheless, the communicative competencies of the agents and their knowledge base can be expected to set limits to their capacity to (a) understand the signals in the network and also the situational meaning in which the network structure resounds, (b) decompose these two dimensions (that is, the information contents of messages and their meaning), and (c) participate in further communication by reflexive restructuration of this relation—between the information contents of messages and their meaning—in follow-up communications. The two systems layers ("consciousness" and "communication") can be considered as reflexively co-evolving (or not!). This is appreciated by Luhmann (1977)—following Parsons (1968, at p. 437)—as "interpenetration."

---

[1] Luhmann, 1984, at p. 226 [1995, at p. 164]: "Die wichtigste Konsequenz dieser Analyse ist: *daß Kommikation nicht direct beobachtet, sondern nur erschlossen warden kann.*"

The (latent) structure and (virtual) dynamics at the network level can be expected to develop non-linearities because of the recursive operation of interactions. The outcomes may hence be counter-intuitive for participants, and unintended consequences of purposeful action can be expected to prevail in a networked environment. Analytical methods like factor analysis were developed precisely for the purpose of revealing these latent structures.[2] However, the focus in these methodological efforts has been on complexity at a specific moment in time, while meaning operates in terms of updates over time. The extension of the factor-analytic design to a dynamic system is methodologically not a *sine cure* because one needs theoretical guidance to distinguish change in the observable variables from change in the latent eigenvectors (Leydesdorff, 1991, 1997).

From this perspective, Luhmann's most important contribution has been his creative combination of Maturana and Varela's (1980) biologically inspired theory of *autopoiesis* (or self-organization) with Parsons's structural-functionalism. Parsons theorized the latent functions for the case of the social system. In Parsons's so-called four-function paradigm, however, the functions were limited to four or nested multiples of four for analytical reasons (Parsons, 1970).[3] Merton (e.g., 1973) noted that norms and counter-norms historically generate ambivalences about their functionality in social systems and that functionalities may therefore change over time (Mittrof, 1974). Merton abandoned Parsons's rigid scheme in favor of a focus on historical change.

---

[2] A network is constructed in terms of relations ("links"), but the resulting architecture of the network also positions the nodes and links in it. Factor analysis of the matrix representation of a network enables us to distinguish the main axes of such architecture.

[3] The number four was originated by cross-tabling two dichotomies: internal versus external and instrumental versus consummatory (Adriaansens, 1976, at pp. 151f.; Parsons, 1970, at p. 31).

In Luhmann's theory the functions are no longer given, but culturally constructed. The processing of meaning in inter-human communication is considered recursive: some meanings can be considered as more meaningful than others. Knowledge, for example, can be considered as a meaning which makes a difference. Meaning can make a difference if the knowledge is codified, for example, as discursive knowledge. However, the dynamics are more general: in the court room, for example, the alleged intentions of the accused can mark the difference between murder in the first or second degree, given the legal code which uses these categories for its coherence and organization. Thus, the development of the communication of meaning in terms of codifications determines the complexity which can be handled by a social system.

Knowledge can be codified, but it codifies in turn the underlying meaning and information. In other words, codification is a recursive operation. At the very bottom, there is only uncertainty or probabilistic entropy (Shannon, 1948). When uncertainty is positioned in a network of relations, this position can become meaningful. However, meaning is provided by a system which itself is updated by *positioning* the relational information. Furthermore, meaning is provided from the perspective of hindsight, that is, against the arrow of time, *and* with reference to future possible meanings. Three selections upon the prevailing uncertainty are thus relevant: (1) the selection at each moment of time, which structures the complex system as a configuration with latent dimensions; (2) the selection over time by the system, which potentially stabilizes (or de-stabilizes) the system; (3) a selection among other possible meanings. Luhmann makes reference to Husserl's (1913, 1929) notion of a horizon of possible meanings for the specification of this last selection (cf. Luhmann, 1986b).

The horizon of meanings can be considered as next-order or global from the perspective of the historical stabilizations and consequent institutionalizations of meaning (Grant, 2000). Parsons (1963a and b) introduced "symbolic generalization" of codes of communication to explain the sometimes binding character of collective action. While the first selection can be considered as subsymbolic and the second as symbolic—since the information is not only positioned, but the information and its position are also acknowledged by the positioning system—the third selection from a horizon of meanings is based on symbolic *generalization*. Using money as the model of an exchange medium, Parsons generalized this concept of media to include power and influence as other exchange media that can be generalized symbolically.

Luhmann (1975) elaborated on these concepts by considering power, etc., as codes of communication, each of which can be generalized in different directions. The functional differentiation in the codes of communication could thus be associated with the factor-analytic model of eigenvectors. Under the condition of modernity exchange media can develop according to their own logic, and thus science, politics, the economy, affection, etc., can further develop their own specific codes of communication along analytically orthogonal axes.

The codes operate also in parallel, and thus generate co-variations. For example, everything can be assessed in terms of its economic value or its esthetic beauty, and these assessments can no longer be expected to correspond because they are coded differently. When the codes of communication are functionally differentiated, for example, among scientific communications, religious communications, economic transactions, or political power relations, more complexity can be processed by the communication system as a whole. However, because of this functional differentiation among the codes of communication, the "whole" tends to disappear: the more complexity can be processed, the more transformative the system becomes. Unlike living systems, the social system of communications does not need to be integrated into an identity performing a life-cycle, The social system can be expected to remain differentiated and therefore transformative—using differentiated fluxes along nearly orthogonal axes—on the historical organizations of meanings.

When the functional differentiation prevails, the latent dimensions (eigenvectors) can increasingly take over control functions from the observable relations (vectors). In other words, the feedback arrows (in Figure 1) from the differentiated self-organization of the communication fluxes towards the historically contingent organization of the system can become more important than the bottom-up arrow from the historical basis towards the transformative dynamics of meaning processing.

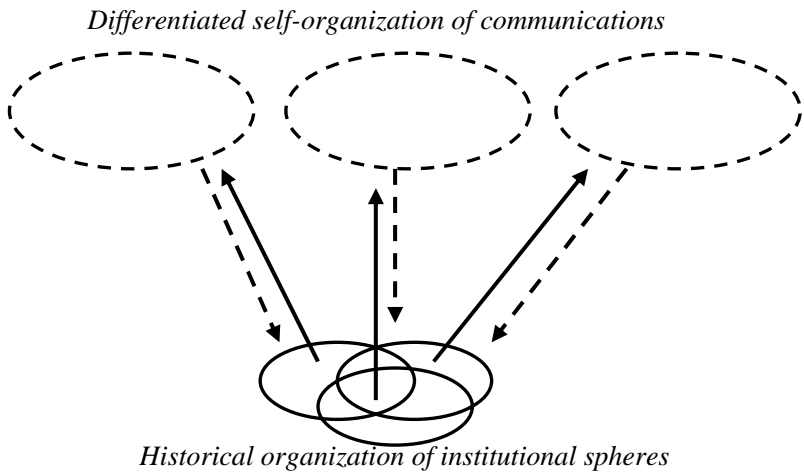

*Differentiated self-organization of communications*

*Historical organization of institutional spheres*

**Figure 1**: The model of differentiation in the communication and integration in historical organizations

The differentiation among the codes of communication can be considered as a historical variable. The more the system is differentiated, the more complexity it can process. However, the functionally differentiated subsystems remain analytical constructs which co-vary in historical events. For example, a decision to organize a school or a household in one way or another has different meanings in terms of this entity's earning power, legal status, etc.

Historical events can be considered as instantiations of interactions among the functional subsystems. In Giddens's (1984) language, these subsystems provide the "rules and resources" which structure society historically. In other words, the subsystems are "operationally closed" *in the analytical model,* but not at the level of historical phenomena: the subsystems can be considered as abstract dimensions ("genomena") in which events can be provided with meaning in accordance with the codes of communication. The latter are organized in and reorganized by the historical instantiations. The differently coded meanings hang together and are therefore specifically organized in the historical events to which they provide meaning.

This historical anchoring of the systems of communication in people, organizations, and institutions makes the system at the same time transformative and historically contingent. Differentiation and integration are two sides of the same coin. However, the "coins" are no longer given as hardware, but as operations that develop over time. Thus, the model becomes algorithmic instead of geometrical. In my opinion, this model is stronger than any of the preceding ones. It combines the strength of the factor-model of so-called "eigenvectors" with the dynamic perspective of the theory of autopoiesis.

The factors are evolutionarily relevant insofar as they serve as functions to the autopoiesis of the system. Since they are operational, they are not given as observables. Reflexively, one can hypothesize them in order to carry the explanation. This provides the factor model with an empirical operationalization: communications can be measured (for example, in terms of clusters of words or messages) and analyzed in terms of the eigenvectors of a matrix, and one can try to identify the eigenvectors as indicators of functionally different codes of communication (Leydesdorff & Hellsten, 2005). The model makes us aware that both the codes of communication and their content can be expected to change over time.

Unfortunately, the later Luhmann no longer offered his theory as an empirical program, but as a universal theory including a different epistemology (Luhmann, 1997). In the remainder of this paper, I focus on three issues in which the elaboration of an empirical program of research can be considered incompatible with Luhmann's universal theory. These are the concepts of information, observation, and language.

**Information**

Luhmann (1984) followed Gregory Bateson (1972, at p. 462) in defining information as "a difference which makes a difference." Bateson defined this as "negentropy" at the level of general systems theory. Luhmann applies this concept of information to the social system which processes meaning. In other words, the difference makes a difference for the differences in meaning. Let us label the differences with subscripts and formulate as follows: the difference$_1$ makes a difference$_2$ for the differences$_3$ in meaning.

Differences$_1$ can be considered as sources of uncertainty or Shannon-type information. This flux of information is discarded by Luhmann (e.g., 1984, at p. 290; 1995, at p. 213) as "erratic:" it is noise or disturbance but not yet meaningful information. By making a difference$_2$ for the meaning processing, the uncertainty can be selected as meaningful information. As Luhmann put it (1995, at p. 67): "By information we mean *an event that selects systems states*." Because the system is considered as autopoietic, it self-selects these system states. It can only absorb information that is already meaningful within the system (difference$_3$). All other uncertainty is deselected as noise (Grant, 2007).

The empirical researcher can analyze both differences$_1$ and differences$_2$, if the research design is reflexive enough in distinguishing between what differences may mean from the perspective of participants and analysts (Mulkay *et al.*, 1983; Gilbert & Mulkay, 1984; Leydesdorff, 2006). For example, in the information sciences distributions of citations have been used as indicators of codification (Fujigaki, 1998), and words and co-occurrences of words as indicators of the change in the meaning of words in contexts (Callon *et al.*, 1986; Leydesdorff & Hellsten, 2005). Differences$_3$ in the internal meaning processing within systems, however, can only be measured if the mechanisms of this processing are properly specified. In other words, systems theory is needed for the specification of hypotheses and heuristics.

The hypotheses can serve to explain the observable (changes in) distributions of communications that would result from the theoretically specified processing. However, the Luhmannian programme discarded this effort as the explanation of noise (differences$_1$) in favor of a focus on the relations between differences$_2$ and differences$_3$. The coupling with empirical differences$_1$ is not theoretically appreciated; uncertainty and error do not have to be specified. Empirical uncertainty is denied any status other than that of an external referent (Luhmann, 1990b).

The systems are considered as operationally closed, not only in the model, but also in reality. In my opinion, this resembles a biological model in which one assumes that the underlying system is no longer accessible for the next-order one because of operational closure. This closure can be considered as a membrane. Human languages, however, enable us to reconstruct both the meaning and the empirical information contained in messages (Leydesdorff, 2000). In other words, the closure in language remains provisional. I come back to this plastic ("infrareflexive"; Latour, 1988) function of language in a later section, but let me first explain Maturana's biologically inspired model in more detail.

Maturana (2000, at p. 462f.) noted that "a recursion takes place whenever a circular or cyclical process is coupled to a linear one, that is, when a circular or cyclical process is applied to the consequences (linear relational displacement) of its previous application. When there is recursion, a new operational domain appears." Using this model, the domain of meaning processing in social relations appears as a self-referential and therefore potentially autopoietic domain on the basis of a linear flux of information. The information can be provided with meaning by reflexive agency. This reflexive agency ("consciousness" in Luhmann's wordings) is embedded in social relations. Interaction terms among the agents and the differentiated codifications (operationalised above as eigenvectors of the network) can drive this system into a self-organizing mode. The symbolically generalized media develop as specifications of language.

The languaging among the agents and the next-order symbolically generalized communications (e.g., economic transactions) are measurable as fluxes of information. These can be analyzed statistically (for example, using entropy statistics). One may even be able to analyze what part of this information is considered by reflexive agents ("consciousness") and the communication systems under study as meaningful information and therefore used in the meaning processing of these systems. However, Luhmann's theory uncoupled itself from this information processing by focusing *exclusively* on the circulation of meaning.

**Observation**

The communication of meaning leaves footprints in the empirical world because the two contingencies of information processing and meaning processing are coupled, for example, in language. Although the explanation of empirical contingencies can be considered as outside the scope of Luhmann's theory, the word "observation" nevertheless plays a very important role in this theory. However, the concept of "observation" is provided with a meaning completely different from its use in empirical research, where observations inform expectations.

Following Spencer Brown (1963), "observation" is defined by Luhmann as a distinction which is followed by a designation. In sociological discourse, however, the specification of a distinction followed by a designation generates only an observational category. One can specify an expectation for the value of this category, but one still needs to proceed to measurement in order to test expectations against observations. By calling expectations observations, Luhmann's theorizing is no longer in need of measurements.

This abstract notion of observation became increasingly central to Luhmann's theory during the 1990s to the extent that it became—according to many (e.g., Fuchs, 2004; Baecker, 2005)—more foundational than concepts like meaning and communication in earlier work. I'll provide the quotations below, but let me first mention Hans Ulrich Gumbrecht (2006, at p. 700) who distinguished three stages in Luhmann's intellectual carreer: first, the stage in which meaning was conceptualized as the central concept for sociological theorizing (in discussions with Parsons and Habermas, who focused on "action" and "communicative action," respectively) (Luhmann, 1971); secondly, the absorption of the theory of autopoiesis mainly during the 1980s (Luhmann, 1984); and thirdly, "in the last years of his life the focus was on the concept of observation and on the emergence of levels of meta-observation" (e.g., Luhmann, 1993). Let's call these theoretical programs Luhmann$_1$, Luhmann$_2$, and Luhmann$_3$, respectively.

In my opinion, Luhmann$_3$'s "observation" is based on a misinterpretation of Spencer Brown's (1963) book *Laws of Form.* This book provides a mathematical theory about forms and marks. A central point is that an indication can be shaped by a distinction followed by a designation. At the very last page (p. 76), the author states that "an observer, since he distinguishes the space he occupies, is also a mark. […] We see now that the first distinction, the mark, and the observer are not only interchangeable, but *in the form*, identical." (italics added -LL) The formal analogy, however, does not preclude that the two referents are substantively different. Spencer Brown's concept is a mathematical one, and observers can be specified biologically (Varela *et al*., 1991), psychologically (Edelman, 1989), and also, but differently, in terms of sociological theorizing (Giddens, 1976; Mulkay *et al*., 1983; Leydesdorff, 2006).

Luhmann$_2$ always emphasized the importance of the specification of a system of reference when defining concepts because he wished to distinguish social systems of communication from "consciousness systems," and these latter two systems from living systems (Luhmann, 1986a). The first distinction clarifies, for example, when philosophers discuss the concept of "freedom" in general. A systems theoretician can argue that "freedom" means something different for an individual than the institutionalization of civil liberties at the level of society.

Similarly, a distinction followed by a designation provides us only with an observational category at the level of the social system or in a scientific discourse. One needs to proceed to measurement to provide the box with a value; this requires observations by reflexive agents. Furthermore, at the level of a social system one expects a distribution of possible observations containing an uncertainty (and an error term, which can sometimes be reduced by further observational reports).

Paradoxically, Luhmann (1984, 1990a) initially resisted defining the "observer" as a central category in his theory, against a dominant tendency to do so in systems theory. For example, in his article entitled "Cybernetics of Cybernetics," Von Foerster (1979) countered Maturana's (1969) Theorem Number One that "*Anything said is said by an observer*" with his own Corollary Number One: "*Anything said is said to an observer.*" Luhmann counter-argued in *Die Wissenschaft der Gesellschaft* (1990a, at p. 14) as follows:

> As a result of a long, but unambiguous tradition of attributing knowledge to human beings, one can see a certain idealization of the observer as a complex of measurements and calculations. This is notably the case for modern physics in which one is more reflexive on the effects of the measurement instruments than on the effects of the human operators. Therefore, it would almost be possible to abstract from the interpretation of an observer as a subject and discuss only "to observe" or "observations." Such careful distinctions, however, do not solve the issue because one has "nach wie vor" only one possibility for the identification of an observer and that is as a human being.

On the next pages, Luhmann$_2$ preferred to ground his theory on "double contingency." This is indeed an important move, because the second layer of the double contingency is based not on contingent observations but on the expectation that *Alter* entertains expectations comparable to *Ego's* expectations (Vanderstraete, 2002; Parsons, 1951). On the one hand, the processing of meaning thus deeply anchored in expectations and intentions. Husserl's (1929) quest for whether a intersubjective intentionality can be considered as a monad different from subjectivity was thus brought back on stage and sociologically operationalized (Luhmann, 1986b). On the other hand, the specification of expectations about expectations provides us with options for empirical research in the sociological domain: which expectations are entertained by whom, and why (Berger & Luckman, 1966)? How are expectations specified and codified, for example, in discourse?

During the 1990s, Luhmann$_3$ became ambivalent about the relations between intentions, expectations, and observations. In his *chef-d'oeuvre* of 1997, for example, Husserl and the notion of intentions were completely backgrounded. Increasingly, Luhmann accepted during the 1990s Von Foerster's formal notion of an "observation" as the central category in his own theory. Luhmann (1999, at p. 20), for example, formulated this as follows:

> How communication is then possible remains, it is true, an open question for traditional subjectivism. But our answer can now read: through communication, that is to say, through the formation of a system of observation *sui generis*, that is through the formation of a social system.

However, Spencer Brown had emphasized that the analogy was only *formal*. As noted above, the social system is not an observer other than metaphorically. At the level of an interaction system among observers, a distinction plus a designation cannot yet generate an observation, but only an observational category. The system of social coordination has no sense organs to perform observations; scientific discourse, for example, can only process observational reports. Observations are contextual to the mechanisms of social coordination; communication can provide observational reports with meaning.

In a second step, the confusion got worse. The second-order "observer," that is, the one who is reflexive on the first-order one in the double contingency and thus constitutive of the social system as an inter-human communication system, is turned into a first-order one when Luhmann$_3$ (1999, at p. 20) formulated that "the second-order observer, mind you, is a first-order observer as well, for he must distinguish and designate the observer he intends to observe." Thus, the observer is at the same time made formal (allegedly with the authority of Spencer Brown) and naturalized as a first-order observer ("he").

In my opinion, the sociological enterprise begins with reflexivity about the analytical differences between the observational reports of participant observers and analytical observers. Giddens (1976) elaborated the potential difference in positions between observers and participants in terms of a "double hermeneutics," and Geertz (1973) distinguished "emic" from "etic." Naturalistic observations are out of the question because the participants' and the analysts' roles can then become analytically confused. When the first-order observation of a participant is equated with the second-order one of a reflexive analyst without sufficient reflexivity about the changes in position specified in a research design, the sociological research program can be expected to degenerate (Leydesdorff, 2006).

**Language**

In a follow-up to his debate with Luhmann$_1$ in 1971, Habermas (1986) acknowledged that Luhmann$_2$ had made important steps: the subject as the centre of reflection in traditional philosophy was replaced with a system's notion, which objectifies Husserl's notion of an intersubjective monad. Habermas (1986, at p. 385) praised Luhmann$_2$ as a sociologist:

> As Luhmann's astonishing job of translation demonstrates, this language can be so flexibly adapted and expanded that it yields novel, not merely objectivating but objectivistic descriptions even of subtle phenomena of the lifeworld.

According to Habermas, this objectivation replaces meta-physics with a meta-biology because meaning-processing in the social system is prelinguistically defined as a referential context of actualizable possibilities that is related to the intentionality of experience and action (*Ibid.*, p.369). In other words, the horizon of meanings is naturalized as transcendentally given. (In this respect, Luhmann followed Husserl, for whom the meditation precedes the discourse.) According to Habermas, however, this sociological theory—which externalized both consciousness and language—had unfortunately led to a concept of monadically encapsulated consciousness systems (à la Robinson Crusoe) who live in administrations with mailboxes for incoming and outgoing communications.

Habermas's alternative of "linguistically generated intersubjectivity," in my opinion, is not incompatible with the systems-theoretical approach, but the linguistic medium of communication has then to be specified as more complex than telephone lines. Künzler (1987, at p. 323) noted that "contrary to Parsons who used *linguistic* models for code of communication, Luhmann wished to derive his media-theoretical concept of code from the model of *genetic* code." Luhmann understands "code" as a rule for duplication. Against Parsons, Luhmann emphasized that the symbolically generalized media of communication are not a specialization of language, but a supplement to it.

According to Künzler (1987, at p. 331), Luhmann$_2$'s model is meta-biological because language "is considered as a disturbing element which cannot be eliminated" from systems theory or its subtheories. The evolutionary perspective of explaining human language and its further development into symbolically generalized codes of communication in terms of exchanges of information is not developed. The function of language and symbolically generalized media of communication as supra-individual coordination mechanisms was not elaborated, although the materials were available in the writings of Parsons and Habermas. How can one understand language and codification in terms of the model of autopoiesis at the level of the social system?

Let me follow Luhmann (2002, at p. 175) that human language can be considered as an evolutionary achievement. The evolutionary achievement can be specified using communication theory (Leydesdorff, 2000, 2001): language allows us to communicate using two channels for a communication at the same time without fixing the interface between them. Cells in a neural net can only fire at different frequencies and thus communicate (Shannon-type) information. Upon reception, this information can be converted into meaningful information by discarding the noise. The meaningful information can again be communicated in a next-order domain (Maturana, 1978). However, language enables us to communicate both uncertainty (that is, Shannon-type information) and meaning. The meaning is not in the message, but it can be reconstructed from the message because the message is expected to contain both information and meaning. The meaning is imprinted as meaningful information which can be reconstructed by a receiver from the information received. The meaning is not given in the information. However, one is able to distinguish between the information content of a message and its meaning, and bring this distinction again into the communication.

Not all information is meaningful, and meaning can also be communicated without using language. In methodological terms, one can think of codes of communication as one more (that is, a third) degree of freedom in differentiated communications. The codes enable us to focus on specific selections. From an evolutionary perspective, selections can be stabilized by codification if they are functional for the reproduction of the system. The codes of communication thus add a third dimension to the communication. Some information is more meaningful than others, but the meaning at the supra-individual level is decided in a social system in accordance with the prevailing codes of communication.

When the coding prevails, a jargon is formed which closes the group of potential communicators, or the communication can even go on without using words; two (of the three) dimensions are already sufficient for meaningful communication. In inter-human communications, however, the coding remains embedded in the linguistic processing as a next-order recursion, just as the linguistic exchange remains embedded as a recursion in the information exchanges. Once in place, however, there is no longer reason to prioritize one layer over another: one can leave this open as an empirical question.

For example, on the market place one can pay the price of a commodity with money (coins and banknotes) without further communication in language. In another context one may have to negotiate about the price in language. More abstract payment by credit cards requires a signature or an equivalent exchange of linguistic symbols. Using another symbolically generalized medium of communication, one may not need many words to express one's affection in an emotional relation, but in most cases a linguistic communication is also needed (e.g., "I love you."). Roses do not always do the job without words.☺ In other words, one obtains an empirical field of research questions about how codes of communication operate and sometimes make linguistic communications superfluous. The inter-human communication system has three dimensions available for processing this complexity: uncertainty, mediation, and symbolization.

Using Maturana's (2000) definition of a recursion (provided above), we are now able to specify the differences between biological systems, psychological systems, and social systems of communication in these terms. A neuron fires or does not fire: (Shannon-type) information is transmitted. The channel is one-dimensional; both at the sending and the receiving ends an embedded observer is needed. In other words, the brain cells are structurally coupled to the neural net. If a recursion occurs, however, another domain can be shaped endogenously. This domain is operationally closed as a next-order cycle on top of the information flow. The non-linear cycle allows the system to handle more complexity, but in the biological system each level has its own mechanism and medium. If the levels were allowed to penetrate into each other, the biological systems under study would be at risk of dying. This is most obvious in the blood-brain barrier.

Human language is an evolutionary achievement. It is not just an extension of the biological domain with another recursion. The evolutionary achievement, in my opinion, is that two channels of the communication are available, and this distinction is no longer hardwired. This allows us to reconstruct the information received in terms of uncertainty (Shannon-type information) and meaningful information. One can expect that a message in language is provided with meaning by a sender. The meaning can be reconstructed from the message by a receiver because the message contains meaningful information in addition to Shannon-type information. The coupling between the carriers of the communication and the communication is not only structural, but also operational because of this interface between meaningful information and uncertainty *within* the message. A system which operates with *messages* is different from a (biological) system which operates in terms of signals.

The volatile operation of providing meaning to the messages is recursive at the level of the social system, but not yet at the level of individual consciousness because of the latter's embodiment. On top of the information flow and the flow of meaningful information, a next-order layer at the level of the social system can be shaped along the eigenvectors of the communication network which no longer rests on embodiment—and hence on a life-cycle—but on codifications.

While these eigenvectors are spanned orthogonally, the system can increasingly be expected to drift towards functional differentiation in terms of the codes of communication. However, the eigenvectors are analytical constructs which remain latent for the participating communicators. They operate "behind our backs" (Marx). As the codes are functionally differentiated, they can accelerate the communication of meaning by orders of magnitude. The social system thus gains a degree of freedom for processing complexity.

Unlike the hardwired dividing lines among operational domains in biological systems, the communication of uncertainty, meaningful information, and meaning remains permeable for one another. These distinctions are not hardwired, but culturally constructed. The codes of communication remain embedded in language, but they can operate beyond language. A reflexive agent can always prefer a lower acceleration: if uncertainty arises about the quality of the communication at the symbolic level, explanation in language is needed. One may wish to go back to the first acceleration by writing everything out so that the distinction between the information content and the meaning of the message can be more objictified than in interactions. The communication can then even be faxed over a telephone line (that is, in terms of Shannon-type information).

The three channels are contained in the evolutionary achievement of inter-human language. Languaging is different from barking among dogs or signaling among delphins. Not only information is conveyed, but also meaning. This can be done because the information has a position in a semantic field. When the three layers are conceptualized as horizontally super-imposed in a hierarchy, the complexity cannot yet be unfolded. The orthogonalization of the codes of communication under the condition of modernity dissolved the idea of control. Self-organization in the communication of meaning becomes increasingly possible when the codes of communication can be appreciated as an additional degree of freedom.

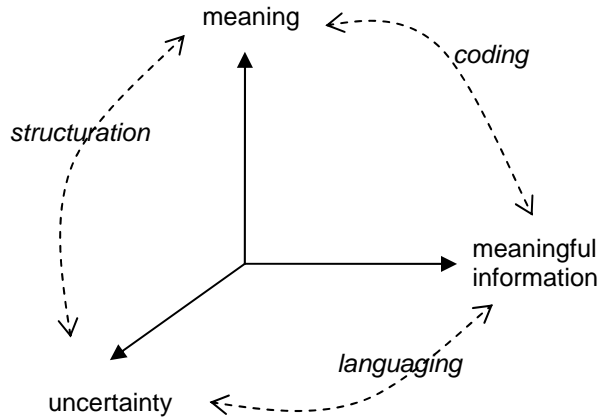

**Figure 2**: Three dimensions of communication

Figure 2 shows this unfolding of the capacity of interhuman communications in terms of relevant dimensions. The image would become even one step more complex if one could account for *functional differentiation* of the codes of communication. The symbolically generalized codes then become feedbacks on all the axes and planes between them. For example, the structuration becomes different depending on which symbolically generalized code prevails in a given instance. All codes resound, but in specific instances some are suppressed more than others.

Herbert Simon (1962) conjectured that all systems can be considered as composed from alphabets. Perhaps some 20+ codes are involved, perhaps less. The combinatorial possibilities are immense because the possible languages are multiplied by this additional dimension of next-order codifications. The codes resound in all inter-human communication, whether the communication is linguistic or not. Note that Luhmann's premise that everything that happens between human beings can be considered as communication and thus be analyzed in these terms requires an empirical theory of communication complex enough to explain the non-trivial turns that the social system can take. Monkeys and dogs are able to understand symbolic gestures, but they don't have the richness of a language available. Three subdynamics are needed to generate the full scope of non-linear possibilities (May & Leonard, 1975).

For example, in a scientific discourse one can exchange data without much wording, argue with words, and/or invoke the code of communication which Luhmann characterized as true or false. By considering the codes as spanning dimensions one obtains all the grey shades of communication and therefore room for empirical analysis. A richer dimensionality enables us to bring together what has been remote, that is, by providing it with a new meaning in terms of a *translation* of one geometrical representation into another.

Using geometrical metaphors, representations in language can be conceptualized as either instantiations (at one moment of time) or trajectories (over time). In the case of an instantiation one explains the complexity in the aggregate at a certain moment and in the case of describing a trajectory the time axis is used for structuring the narrative. As Giddens (1984) noted, one of the available dimensions is bracketed in either case. Luhmann (1984) used the metaphor of a blind spot implied in a theoretical appreciation, more generally. One needs to select one background or another to stabilize a perspective. Because Luhmann did not develop this concept of a *translation* among geometrical metaphors, he considered different perspectives mainly as paradoxes to be resolved historically instead of analytically (Law, 1986).

By focusing on translations as changes in the specification of uncertainty, an external super-observer can be reintegrated into the linguistic domain among participant-observers only if the medium allows for reflexive communication. Whereas a biological function has to be reproduced (following the pattern given in the life-cycle), it remains dependent upon the reception among the participant/observers whether and how a social function is shaped. In a high-culture, roles were pre-ordained and function tended to be equated with meaning while this social system was stratified. In this configuration, the super-system is still identifiable (e.g., as the King). When the cosmological order is broken, roles may be reflexively acknowledged or not, and the meaning of a communication can increasingly be distinguished from its function.

**Conclusions**

The bridge between theorizing and empirical research provides a crucial step because only operationalizable theories can be methodologically informed and controlled. Luhmann[2] acknowledged this, for example, when he formulated: "In order to validate the binary code of true and false, one needs programs of a different type. We call these programs *Methods*." (Luhmann, 1990a, at p. 413). On the next page he specified that "one can consider it the task of a methodology to keep track of the difference between the first and second order." Note that this reflection is not so far removed from Giddens's (1976) "double hermeneutics" except for Giddens's awareness that this distinction cannot be made without uncertainty, and therefore reflexivity.

Luhmann believed in black and white (true or false as in logic)—"um zur Entscheidung für den einen statt den anderen Wahrheitswert zu kommen"—while empirical research requires the expression of uncertainties and the testing of significance at a probability level. Luhmann (1990a, at p. 176) himself italicized the sentence "*Truth itself is not relative*." In my opinion, this claim is inconsistent with the idea that truth can be considered as a symbolically generalized code of communication. If one accepts that truth is the code of communication in scientific discourse, then it should operate as a latent and therefore uncertain dimension of this discourse.

The starting point of my perspective is the uncertainty which prevails and its expression in bits of information. Shannon's (1948) mathematical theory of *communication* can be elaborated into an information calculus which includes the time dimension (Bar-Hillel, 1955; Theil, 1972; Brooks & Wiley, 1986; Leydesdorff, 1995 and 2001). Only a calculus—beyond a logic—can account for time. We are in the fortunate situation of having both this methodological apparatus available and the rich theorizing from Luhmann$_2$. However, the empirically interested social scientist should resist the tendency in Luhmann$_3$ to develop a grandiose theory in which everything is sufficiently explained insofar as it can be distinguished and designated.

The naturalization and generalization of "observation" as a foundational category by Luhmann$_3$ generated a meta-biological research program. This meta-biological research program is anti-humanistic and anti-sociological because reflexivity in inter-human communication and codification is no longer sufficiently appreciated (Husserl, 1929). The "double contingency" between expectations and observations is resolved in favor of the latter. Consequently, observation is not considered as a construct of a (potentially scientific) discourse and uncertainty in the (double!) contingencies can no longer be specified.

**Acknowledgement**
I am grateful for comments of Peter Fuchs, Rob Hagendijk, Franz Hoegl, and Jörg Räwl on previous drafts.